\documentstyle[prd,aps,multicol]{revtex}

\draft

\title{Strangelet spectra from type II supernovae}

\author{ H.Vucetich$^{1}$\thanks{Member of CONICET} and J.E.Horvath$^{2}$}

\address{$^{1}$ Facultad de Ciencias Astron\'omicas e Geof\'{\i}sicas, 
Universidad Nacional de La Plata,\\ Paseo del Bosque S/N, (1900) La Plata, 
Argentina.}

\address{$^{2}$Instituto Astron\^omico e Geof\'{\i}sico, Universidade
de S\~ao Paulo,\\ Av. Miguel St\'efano 4200, \'Agua Funda,
04301-904, S\~ao Paulo, Brasil. }

\date{\today}

\begin{document}

\maketitle

\begin{abstract}
\widetext
We study in this work the fate of strangelets injected as a contamination 
in the tail of a "strange matter-driven" supernova shock. A simple model 
for the fragmentation and braking of the strangelets when they pass 
through the expanding oxygen shell is presented and solved to understand 
the reprocessing of this component. We find that the escaping spectrum is 
a scaled-down version of the one injected at the base of the oxygen shell. 
The supernova source is likely to produce low-energy particles of 
$A \, \sim \, 100-1000$ quite independently of the initial conditions. 
However, it is difficult that ultrarrelativistic strangelets (such as the 
hypothetical Centauro primaries) can have an origin in those explosive 
events. 

\end{abstract}

\twocolumn

\section{Introduction}

More than a decade ago a celebrated paper by Witten \cite{Witten} (see
also the previous works by Bodmer \cite{Bodmer} and Chin and Kerman
\cite{ChinKer}) suggested that we may have been overlooking the true
ground state of hadronic interactions.  According to this work, {\it
strange matter} (cold catalysed quark matter nearly symmetric in $u,d$
and $s$ flavors) would offer a possible form of bypassing the
limitations imposed by the Pauli principle in ordinary matter because
of the existence of a third (strange) Fermi sea to share the energy of
the system. Then, the energy per baryonic number unit would be smaller
when compared to two-flavor quark matter and if this reduction is
large enough, strange matter created after a weak-interaction
time-scale may be the lowest energy state. Following this suggestion,
which was essentially based on a bulk Fermi gas picture, the
properties of strange matter and the droplet version, termed
strangelets, were investigated \cite{FahrJaff,BerJaff,Madsen95}.
Particular attention has been paid to the possible shell structure in
the few-quark strangelets \cite{TakaBoyd,Michel88,GilJaff}, which are
the ones expected to show up in heavy-ion collision experiments
\cite{Pretzl}. A great deal of papers also explored astrophysical
scenarios which could render a non-zero ISM abundance of strangelets
(which would cause all neutron stars to become strange stars)
\cite{Witten,BH89,Glend90,RuGla,MTH96}, the latter point being
specially important because of the criticisms \cite{Alpar87,Madsen88}
raised to the SQM hypothesis based on pulsar glitch observations, to
which the basic strange star models \cite{SS} have no
reasonable explanation to offer (see Refs. \cite{glitch-lit} for more
elaborated stellar models possibly containing an explanation of that
data).

Concerning the experimental detection at the Earth, it is interesting
to note that, even before the official "invention" of SQM by Witten,
some cosmic ray events \cite{Lattes} were tentatively associated with
quark blobs primaries \cite{BjoMcL}. A summary of the reported SQM
candidates in several experiments is shown in Table 1.


In Refs. \cite{BH89,Glend90,MTH96,Saito,BoySa} aspects of SQM
production have been investigated and discussed. One particularly
puzzling aspect of all candidates is their relatively low baryonic
number $A$. Calculations indicate \cite{FahrJaff,BerJaff,Madsen95}
that SQM tends to be more tightly bound for increasing $A$, so that
injection of favored SQM fragments $\geq \, 10^{5} \, amu$ in any
astrophysical event would require substantial reprocessing to get down
to $\sim \, 10^{2} \, amu$ strangelets. The authors of
Ref. \cite{BoySa} have estimated the reprocessing time-scales by using
model spallation cross sections of fragments of $\Delta \, amu$ with
$H$ and $O$ of the form

\begin{equation}
\sigma (\Delta) \, = \, \sigma_{0} \,  
{\biggl( {m \over{m_{0}}} \biggr)}^{2/3} \, 
\exp ( - \Delta / \Delta_{0}) \; \, ;     \label{FragSigma}
\end{equation} 
where $\Delta_{0}$ is the preferred emitted cluster; and found that 
reprocessing is very ineffective, unless the strangelets can 
pass through a very dense oxygen shell. Although this situation is very 
unlikely for the popular double-degenerate coalescence scenario
\cite{Witten,Glend90}; 
it is precisely the situation expected in strange matter-driven
supernovae scenarios \cite{BHPRL,Deton}. In the latter a second shock
arises because of the exothermic transition $n \, \rightarrow \, uds
\, + \, energy$ and is expected to carry a contamination of
strangelets in the low-velocity tail as a byproduct of turbulent
mixing \cite{BH89,Deton}.  Ejection of the strangelets with $v
\, \sim \, 0.1
\, c$ is one of the possible mechanisms for a non-zero ISM abundance
and thus competes with the coalescence events. The relative
frequencies of both phenomena $\sim \, 10^{-2} \, yr^{-1}$ and $\sim
\, 10^{-4} \, yr^{-1}$ respectively would be enough to identify the
dominating source if the total mass ejected in strangelets could be
calculated. However, since we lack of reliable estimates of the latter
we have tried to infer the mass working backwards from the reported
events [14]. We have obtained $10^{-6} \, M_{\odot}$ and $3 \, \times
\, 10^{-13} \, M_{\odot}$ by normalizing the flux to the events
reported in Ref. \cite{Saito90} and 
Centauros \cite{Lattes} respectively. However, these
results refer strictly to the abundance of SQM primaries with $A \,
\sim \, 100$ measured at the Earth, and therefore the question of the
most likely "isotope" escaping from whatever source to the ISM still
remains.

\section{Spallation of strangelets by oxygen}

Let us address the specific case of SQM supernova ejection. As
explained above, the basic picture postulates of a (mildly relativistic to
non-relativistic) strangelet gas travelling in the tail of the
secondary shock, which encounters the expanding oxygen
shell. According to Ref. \cite{FowHoy}, 
the expansion of the dense oxygen may
be modelled by a number density evolution of the form $n (t) \, = \,
n_{0} \, \exp (-t / \tau_{exp})$ where $\tau_{exp} \, \simeq \, 0.1 \,
s$ is a multiple of the free-fall time-scale appropriate for that
physical situation. The (heavy) strangelets encountering oxygen
targets will loose energy and baryon number, a process that may be
described in the hydrodynamical approximation 
by the following set of coupled equations

\begin{equation}
{d m \over{dt}} \, = \, - \int d\Delta \, n (t) \, \sigma (\Delta) \, v
					\label{mdot}
\end{equation}

\begin{equation}
m \, {d v \over{dt}} \, = \, - C \, {\pi \over{2}} \, 
m_{ox} \, n(t) \, v^{2} \, R_{0}^{2} \, 
{\biggl( {m \over{m_{0}}} \biggr)}^{2/3} \, + \, \dot{m} \, \xi \, v ,
					\label{pdot}
\end{equation}
where $v$ is the velocity of the strangelets relative to the
expanding oxygen shell, $m_{ox}$ is the mass of an 
oxygen nucleus, $R_{0}$ is
the reference radius corresponding to an $A \, = \, 200$ strangelet
(taken to be $5.8 \, fm$) and $C$ is the von Karman
constant. We note that the r.h.s. of eq.(3) has been integrated over 
the fragment distribution and that we have included in the parameter 
$0 \, < \, \xi \, < \, 1$ 
the details of the transfer in the reaction $A \, + \, ^{16}O \, 
\rightarrow \, A' \, + \, ^{16}O' \, + \, \Delta \, + \, energy$. 

Note that we have neglected the absorption of oxygen nuclei by the 
strangelet. These fusion reactions can be crudely described by a geometric 
cross-section times the well-known Gamow factor. A simple calculation 
shows that the quotient of the fusion to the spallation cross-sections is 
$\sim \, \exp(-32 \pi \alpha A^{1/3} c v^{-1})$, where $\alpha$ is 
the fine structure constant and the approximation $Z \, \sim \, 2 A^{1/3}$ 
has been made \cite{Hei}. Absorption is thus suppressed by a large factor 
and is never important in the process we are considering.

Eqs.(\ref{mdot}) and (\ref{pdot}) can be combined into a single
differential equation for the momentum loss which has the solution

\begin{equation}
v \, = \, v_{i} \, {\biggl( {m \over{m_{i}}} \biggr)}^{D - 1} \, \, ,
					\label{v(m)}
\end{equation}
where $v_{i}$ and $m_{i}$ are the initial velocity and mass 
of the strangelet and 
$ D \, = \, (\pi C \, m_{ox} \, R_{0}^{2}) / (2 \Delta_{0} \, \sigma_{0}) 
\, + \, \xi$. 
Going back to eqs. (\ref{mdot})  and
(\ref{pdot}) we obtain the evolution  
of the strangelet mass with time as

\begin{equation}
m \, = \, m_{i} \, {\biggl[ 1 \, - \, ({4 \over{3}} - D) \, 
{\tau_{exp} \over{\tau_{i}}} \, {\Delta_{0} \over{m_{i}}} \,
f(t) \biggr]}^{3 / (4 - 3 D)} \, \, ;
					\label{m(t)}
\end{equation}
with $\tau_{i} \, = \, (n_{0} v_{i} \sigma)^{-1}$ the initial mean-free path 
of the strangelets in the oxygen shell and 
$f(t) \, = \, (1 \, - \, \exp (- t / \tau_{exp}))$.

Finally, we find using eqs.(\ref{v(m)}) and (\ref{m(t)}) 
the kinetic energy of the strangelet 

\begin{equation}
E_{K} \, = \, {\biggl( {m \over{m_{i}}} \biggr)}^{2 D - 1} E_{K,i} \, ; 
					\label{Emin}
\end{equation}
and therefore the total energy of the strangelet $E_{T}$ from 
$E_{T}^{2} \, = \, m^{2} \, + \, E_{K}^{2}$.

\section{regimes of fragmentation}

An inspection of eqs.(4), (5) and (6) reveals that there are different 
regimes of fragmentation separated by the value of the exponent $D$. In 
all the four cases to be discussed below the mass of the strangelet 
decreases (i.e. spallation occurs) irrespective of the specific value 
of $D$; until the available kinetic energy in the center-of-mass frame 
$E_{K}^{c.m.}$ is not enough to strip a chunk whick is bound by an  
amount $\Delta_{0} \, E_{b} / m_{p}$ ($E_{b} \, \sim 10 \, MeV$ is the 
binding energy per baryon number unit in the strangelet). The strangelet 
will survive as long as the latter condition can be reached before the 
mass is driven to zero (or, more precisely, to a value below $m_{min}$ 
corresponding to the minimum stable baryon number of the strange matter 
$A_{min}$). The possible regimes are

\subsection{$0 \, < \, D \, < \, 1/2$}

If $D$ belongs to this range, eqs.(4) and (6) show that both the kinetic 
energy and velocity grow with time. It follows immediately that the 
strangelets can never satisfy the mass freezeout condition imposed on 
$E_{K}^{c.m.}$ because there is always enough energy available to power 
the spallations. Thus, the particles in this regime evaporate completely 
and do not constitute an astrophysically interesting case.

\subsection{$1/2 \, < \, D \, < \, 1$}

In this case the kinetic energy decreases although the velocity increases 
with time. However, in the center-of-mass frame the kinetic energy  
$E_{K}^{c.m.}$ scales approximately as 

\begin{equation}
E_{K}^{c.m.} \, = \, E_{i,ox} \, 
{\biggl( {m \over{m_{i}}} \biggr)}^{2(D - 1)} \, ,
				   \label{Ecm}
\end{equation}

(with $E_{i,ox}$ the initial kinetic energy of the oxygen) 
and can not decrease either. Therefore the particles also evaporate in this 
regime.

\subsection{$1 \, < \, D \, < \, 4/3$}

Now both the kinetic energy and the velocity of the strangelets 
decrease with time. Spallation proceeds until the point in which 
(see eq.7) the freezout condition is reached

\begin{equation}
E_{i,ox} \, {\biggr( {m \over{m_{i}}} \biggr)}^{2 (D - 1)} \, = \, 
{\Delta_{0} \over{m_{p}}} \, E_{b} \, .
				    \label{cond}
\end{equation}

This regime gives rise to a {\it scaling law} for the mass of the form

\begin{equation}
{m_{F} \over{m_{i}}} \, = \, {\biggl( {\Delta_{0} \over{m_{p}}} 
\, {E_{b} \over{E_{i,ox}}} \biggr)}^{1 / 2 (D - 1)} \, ;
				   \label{scal}
\end{equation}

$m_{F}$ being the final or freezeout mass of the surviving strangelet.

\subsection{$D \, > \, 4/3$}

The situation is essentially the same as in the former point, but now the 
mass decreases slowly than before (see eq.5). This does not lead to any 
substantial change in the picture because the timescale 
$\tau_{i} \, \simeq \, 10^{-16} \, s$ is 
so short that the strangelets would not 
be able to travel large distances before freezeout sets in or the 
oxygen shell expands substantially.

\section{discussion}

From the above expressions the full emerging spectrum in mass and
energy could be computed for a given injection spectrum at the source. Even
without performing a detailed 
computation some general features are apparent. First
of all eqs. (\ref{v(m)}-9) show the dependence of the
results with the parameter $D$, which are dramatically different
depending on its actual value.  Recalling the definition and using 
$C \, = 0.5$ as measured in the case of subsonic spheres in 
laboratory we have found a strict lower bound of 
$D \, \geq \, 0.75$ by setting $\xi \, = 0$. 
Thus, it is quite likely that the physical situation corresponds 
to the third and fourth cases above. An important corollary  
is that the escaping spectrum of strangelets is a scaled-down version  
of the one injected by the secondary shock.
A second major point is that
the stripped fragments may or may not decay depending on the (unknown)
actual value of the minimum stable strangelet $A_{min}$. If $A_{min}
\, \sim \, 10$ these fragments are likely to survive further
spallation because of their non-relativistic character. On the other
hand, they would quickly decay into ordinary hadrons if $A_{min} \,
\sim \, 100$.  In any case the spallation interactions of the ejected
strangelets with oxygen are so frequent in the expanding shell that
the final distribution will be concentrated around the lowest stable
"isotope" and the near the minimum (spallation cutoff) energy,
although it is likely that further braking occurs either in the shell
or in the ISM. For example, encounters of the escaping strangelets
with giant molecular clouds are quite frequent and may help to strip a
few baryon number units.  Finally, we note that is difficult in this
scenario to obtain relativistic strangelets such as the ones required
to fit the primaries of the Centauro events. The natural outcome from
supernovae would be non-relativistic primaries in the range $A \, \sim
\, 100-1000$; which are easily obtained from these events if the
strangelets injected initially by the shock have $10^{5}-10^{8} \,
amu$ for $v_{i} \, \leq \, 0.1 \, c$ (see eq. \ref{m(t)}).
A more realistic treatment of the initial strangelet spectrum requires 
a model of fragmentation of the (initially homogeneous) strange matter 
fluid into strangelets in a turbulent environment \cite{Deton}; 
even though we expect the mass scaling law eq.(9) also to hold 
in this case. This subject will be discussed in a future work.

It is very important to measure the actual flux of candidates to
connect it with the details of the source and, through simulations
like the one in Ref. \cite{MTH96}, to measure the total mass in the
galaxy (for example, a recent experiment seems to have measured a much
lower flux than the one determined in \cite{Saito90}). Further
experimental and theoretical studies would be helpful to address the
relativistic ejection of strangelets and the form of the spectra for
each astrophysical scenario. Work on these subjects is in progress and
will be reported elsewhere.

\acknowledgements

We would like to acknowledge the financial support of the CNPq (Brazil) 
and SECYT (Argentina) through Fellowships awarded to J.E.H. and H.V
respectively. The PROINTER Program of S\~ao Paulo University has provided 
financial support to maintain this collaboration.

\clearpage

\begin{table}[htb]
\caption{SQM event candidates
\label{table1}}
\begin{tabular}{cccc}
Z & A & E/nuc [GeV] & Ref. \\
\tableline
- & 75 & $\sim$ 1000 & [20] \\
20-40 & - & $\geq$ 2.22  & [22] \\
14 & 400 & 0.45 & [23] \\
46 & $\geq$ 1000 & - & [24] \\
\end{tabular}
\end{table}
\end{document}